\documentclass[10pt]{iopart}

\usepackage{color}
\usepackage{graphicx}
\usepackage{graphpap}
\usepackage{harvard}
\usepackage{svg}
\usepackage{hyperref}
\usepackage{lineno}

\usepackage[colorinlistoftodos]{todonotes}

\begin{document}

\title[World--Earth Resilience]{Conceptualizing World--Earth System resilience: Exploring transformation pathways towards a safe and just operating space for humanity}

\author{John M. Anderies$^1$, Wolfram Barfuss$^{2,4}$, 
Jonathan F. Donges$^{2,3}$ Ingo Fetzer$^{3,6}$, Jobst Heitzig$^2$,  Johan Rockstr\"om$^{2,5}$}

\address{$^1$ School of Sustainability and School of Human Evolution and Social Change, Arizona State University, Tempe, AZ 85287}
\address{$^2$ Potsdam Institute for Climate Impact Research (PIK), Member of the Leibniz Association, Telegraphenberg, 11473 Potsdam, Germany}
\address{$^3$ Stockholm Resilience Centre, Stockholm University, 11419 Stockholm, Sweden}
\address{$^4$ University of Tübingen,  72076 Tübingen, Germany}
\address{$^5$ Institute of Earth and Environmental Science-Geoecology, University of Potsdam, Potsdam, Germany}
\address{$^6$ Bolin Centre for Climate Research, University of Stockholm, 10691 Stockholm, Sweden}

\ead{m.anderies@asu.edu}

\begin{abstract}

  We develop a framework within which to conceptualize World-Earth System resilience.  Our notion of World-Earth System resilience emphasizes the need to move beyond the basin of attraction notion of resilience as we are not in a basin we can stay in. We are on a trajectory to a new basin and we have to avoid falling into undesirable basins.  We thus focus on `pathway resilience', i.e. the relative number of paths that allow us to move from the transitional operating space we occupy now as we leave the Holocene basin  to a safe and just operating space in the Anthropocene. We develop a mathematical model to formalize this conceptualization and demonstrate how interactions between earth system resilience  (biophysical processes) and world system resilience (social processes) impact pathway resilience.  Our findings show that building earth system resilience is probably our only chance to reach a safe and just operating space.  We also illustrate the importance of world system dynamics by showing how the notion of fairness coupled with regional inequality affects pathway resilience.

\end{abstract}

\maketitle

\ioptwocol

\section{Introduction}

With humanity's departure from the Holocene wherein humans predominantly adapted to global dynamics and entrance into the Anthropocene wherein we increasingly control them, we have transitioned from being passengers on to piloting `Spaceship Earth' \cite{boulding1966economics,orwell2001road,george1884progress}. As pilots, we become responsible for the life support systems on our `Earth-class' spaceship. Until very recently Spaceship Earth has been running on autopilot regulated by global feedback processes that emerged over time through the interplay between climatic, geo-physical, and biological processes. These processes must necessarily have the capacity to function in spite of variability and natural disturbances and thus have developed some level of resilience in the classic sense of Holling \citeyear{HollingResil}--the ability to absorb and recover from perturbations while maintaining systemic features.  

The capacity of these regulatory feedback networks to provide system resilience is limited. The Planetary Boundaries framework \cite{rockstrom2009safe,steffen2015planetary} makes these limitations explicit and defines, in principle, a safe operating envelope for Spaceship Earth.   The pilots face two challenges: knowing the location of the `default' operating envelope boundaries and understanding how these boundaries change with changing operating conditions.  Pilots typically have an operating manual that provides this information which enables them to better utilize the {\em intrinsic} resilience of their vessel (Earth System resilience). Because we don't  have that luxury, the concept of resilience becomes critical: the art of maintaining life support systems under high levels of uncertainty--flying our Earth-class spaceship without an operating manual. Earth System science is, at its core, the enterprise of uncovering the operating manual. Unfortunately, our capacity to  experiment is quite limited: the time required is enormous and the number of independent copies of Earth is limited. We can't test resilience by transgressing global thresholds and observing how the system behaves in new states and potentially recovers, e.g.  `snow-ball Earth' and the `tropical states' of Earth's past.

With a manual, a crew, and a captain, the `resilience' question would boil down to the competence and risk aversion of the captain and the competence of the crew.  This resilience would encompass

\begin{itemize} 
\item piloting the spaceship so as to avoid shocks (e.g. avoid asteroids, maintain safe speeds, etc.),
\item developing knowledge/skills to quickly repair existing systems if shocks can't be avoided,
\item the capacity to improvise and create new systems when existing systems can't be repaired,
\item and conducting routine maintenance so as not to destroy the ship through usage.
\end{itemize}

The first two elements constitute a key element of specified resilience.   The third element (general resilience) is much more difficult to invest in.  It requires the development of generalized knowledge and process to cope with rare and difficult-to-predict events. The fourth element has been the focus of most environmental policy thus far with a tendency to do just enough to get by. 

Now consider our Earth-class spaceship. There is no captain or crew.  Subgroups of passengers are restricted to certain areas (e.g. the upper or lower decks).  Some groups have access to more ship amenities and those with  less access often support the production of amenities for those with more. As with the Titanic, the impact of shocks is very different for first- and second-class passengers.  There is minimal maintenance of life support systems and, in particular, the waste management subsystem.  Complaints about life support systems are met with agreements that it should be fixed but disagreements about who should pay \cite{donges2017math}. Worse yet, there is no operating manual so no one knows how to effect repairs, or their costs.

To an outside observer, our situation might seem absurd.  If given the choice, many rational passengers on board would disembark. While some extremely wealthy passengers seem to be making an attempt, disembarkation is not realistic.  So what do the passengers do?  One key difference between the spaceship metaphor and our journey is that  there is no destination. Further, the ship's journey is much longer than our lives so the ship becomes our home. `Good piloting' is tantamount to effectively managing life support systems--the `Earth System' (ES)--while ensuring the wellbeing of and preventing critical conflict among the passengers--managing the `World System' (WS). And because we must do this without an operating manual, we must build World--Earth System (WES) resilience (WER).  We must be able to define WER to characterizes how close these critical systems are to breaking down and model it to explore mechanisms that enhance or degrade WER.  In the remainder of the paper, we carefully motivate and define WER and link our conceptualization to the existing literature (Section 2).  We then develop a WE model (Section 3) and analyze it (Section 4).   We conclude with a discussion of how insights from our WER analysis can be incorporated into the climate change policy discourse for timely action in the next decade. 

\section{Foundations of `World--Earth' Resilience}

In the classic  ball-and-cup-type resilience concept,  resilience is a system-level property related to the capacity of a system to maintain its structure and function when impacted by perturbation (a `shock') of core controlling dynamics \cite{walerdata2004}.  The challenge for WER is to be specific about what the system is (resilience of what), and what the `shock' is (to what) \cite{carpmetaphor}.  In dynamical systems, the `of what' is a basin of attraction  maintained by a set of feedbacks. A `shock' is then a process that can affect the system but is not in this set of feedbacks.  For WER, the `of what' might be the basin(s) of attraction that support multi-cellular life, e.g. the glacial-interglacial cyclical attractor driven by celestial mechanics in the Pleistocene.  A second attractor might be the `stabilized' Earth as a continuation of the Holocene, driven not by celestial processes, but by human activities (see Figure \ref{fig:WE_basins} and \citename{steffen2018trajectories} \citeyear{steffen2018trajectories}). A third, again due to human activities, might be  `Hothouse Earth'.  The characteristics of this attractor are not fully understood, but pose risks to persistence of human and other complex life forms, and in the worst case, may lead to the destruction of all life support systems on Spaceship Earth.
 
Classic ball-and-cup resilience may be appropriate for ES resilience as just described. However, applying it in a meaningful way is more difficult for WER because long-term evolution of large-scale complex adaptive social systems (cultural, technological, socio-political, socio-epistemic etc.) is open-ended, and long-run attractors might not even exist.  Further, what may be more important than the nature of long-run attractors is the capacity to pilot a system from one to the other (i.e. realize the Earth System stewardship in Figure \ref{fig:WE_basins}A).  That is, the subtle nature of the stability landscape characterising transient dynamics on shorter time scales (Figure \ref{fig:WE_basins}B) is more meaningful in the context of climate change mitigation and sustainable development.  The  robustness and stability of such desirable development pathways of the WES given shocks and conditional on relevant model uncertainties (initial condition, parameter and structural uncertainties etc.) is our focus here. 

Given our definition of the `of what' (transformation pathways), we are now faced with the `to what' question. If we consider the `Earth-with-life system', then exogenous shocks would originate at the solar system scale or larger --- i.e. solar flares, meteor impacts. However, our capacity to transition from one attractor to another also depends on {\em `endogenous shocks'}, i.e. system behavior that deviates strongly from that expected by actors in the system, generated by the internal, non-linear dynamics of the WES such as a financial crisis, pandemic, conflict, etc. These endogenous dynamics both shape the landscape in Figure \ref{fig:WE_basins}B and generate abrupt, difficult-to-predict variations in development trajectories.  The main contribution of our paper is to introduce WER as pathway resilience, develop a mathematical definition and quantification of this pathway resilience that incorporates the interaction between the ES and WS, and provide an example application using a stylized WES model.

Before turning our attention to these issues, we briefly review the latest developments in this area to place the pathway-based notion of resilience we employ here in the context of the broader literature.

\subsection{Planetary Boundaries and Earth System Resilience}

The notion of planetary boundaries defines a basin of attraction for Holocene-like conditions based on limits to how far essential controlling processes can be altered by human activities without risking to undermine the stability of the global system \cite{rockstrom2009safe}. There is no evidence that modern societies can exist, let alone thrive, in conditions substantially different from the Holocene \cite{steffen2015planetary,steffen2018trajectories} and thus maintaining a Holocene-like basin may be essential for our survival \cite{waters2016anthropocene}. Increasing global populations within the last 50--100 years, from 4 Million 10,000 BCE to 7.2 billion in 2019 \cite{UNpop2015,klein2010long} that have driven excessive use of natural resources, massive destruction of habitats, and release of carbon dioxide into the atmosphere, threaten the stability of this basin.

Planetary boundaries define limits that inevitably need to be respected to support the stability of the Holocene-like  basin.  Each boundary crossing increases the risk of  large-scale abrupt or irreversible environmental changes. Humanity has already crossed four boundaries: biodiversity loss, bio-geo-chemical cycling, land-use change and climate change. Self-reinforcing feedback processes that kick in, e.g. control of the global energy balance,  when boundaries are crossed can lock the planet into a non-reversible, increasingly uninhabitable human trajectory for centuries or millennia.  Rayworth \citeyear{raworth2012safe}  makes the important point that creating a just and livable world on Earth requires certain minimum material flows that set minimum values on the key planetary boundary variables for a given population with a given technology. Combining the upper biophysical boundary with the lower socio-economic boundary creates an annulus (a 2-D doughnut) that defines the safe and just operating space, SJOS. 

\subsection{Resilience and Integrated Assessment Models}

Integrated Assessment Models (IAMs) are a class of WES models that are a  standard tool for climate policy \cite{Keppo_2021,mathias2020grounding,van_Vuuren_2016} and typically consist of some low-complexity ES model driven by some more detailed neoclassical growth model. They may have aggregate ES and WS systems (e.g., \cite{DICE_Nordhaus1993}) or some  spatial decomposition in the ES and some sectoral decomposition in the WS (e.g., \cite{RICE_nordhaus1996regional}). IAMs typically focus on  intertemporal optimization problems with a time horizon between a few decades and a century to derive investment and/or regulation trajectories that either minimize costs under some global emissions constraint \cite{DICE_Nordhaus1993,RICE_nordhaus1996regional}, or that optimize a certain trade-off between these costs and climate-related damages. As such, they cannot be used out of the box to study World--Earth resilience \cite{heitzig2018thought,barfuss2018optimization}. In principle, their core models, if stripped of the surrounding optimization algorithm, could potentially be used to study the climate and economic subsystem's reactions to perturbations in order to partially assess the persistence form of resilience. However, only few IAMs include feedbacks from the Earth subsystem onto the economic subsystem (e.g., \cite{IMAGE_alcamo1996baseline}), or include important nonlinear effects within the Earth subsystem such as positive feedbacks or tipping points, both of which are important on the longer time horizons needed for assessing resilience. Also, IAMs are typically already too complex to provide a deep understanding of the system's behavior.  Still, the model of economic dynamics we employ here is derived from the same building blocks as in typical IAMs.

\subsection{Resilience and World Systems Theory}

Quantitative assessments of resilience typically focus on a particular biophysical system and an exogenous,  human-generated press or pulse disturbance, \cite{andrangeresil,anderiesGBsal,carpenterlakes,Jans2004sunk,scheffercatshifts} or a coupled system wherein the balance of model detail is weighted toward the biophysical system and a specific shock such as a flood  or drought \cite{bertilsson2019urban,scanlon2016enhancing}.  Our focus on a `World--Earth' System is an attempt to strike a balance between biophysical, social, and economic processes. We envision the WES as a network of polities, each with its own endowment of human, human-made, social, and natural capital \cite{barfuss2017sustainable,geier2019physics}. The key premise is that it is the endogenous interactions among these nation states, and not the nation states themselves, that is the primary driver of change. That is, polities are not part of a world, but through their interactions, create a world \cite{wallerstein1979capitalist,wallerstein2004world}.  

Specifically, WS theory revolves around the notion that labor markets do not operate at the local, state, or national scale.  WS are predicated on transnational division of labor and exchange of productions based on international trade agreements built on power asymmetries and that create three sets of countries: core, semi-periphery, and periphery.  Because capital is allowed to move and seek out cheap labor, this WS tends to devolve, at least historically, to the relatively wealthy core extracting wealth from the less wealthy peripheries.  This type of `inequality regime' \cite{piketty2020capital} is an important determinant in the capacity of WES to transform to SJOS, as we shall see.

\subsection{Pathway Resilience and Transient Dynamics}

\begin{figure}[t]
  \newcommand{\xdim}{30}
  \newcommand{\ydim}{40} 
  \setlength{\unitlength}{0.1in}
  \centering
  \begin{picture}(\xdim,\ydim)
     \put(-2,0){\includegraphics[width=0.45\textwidth]{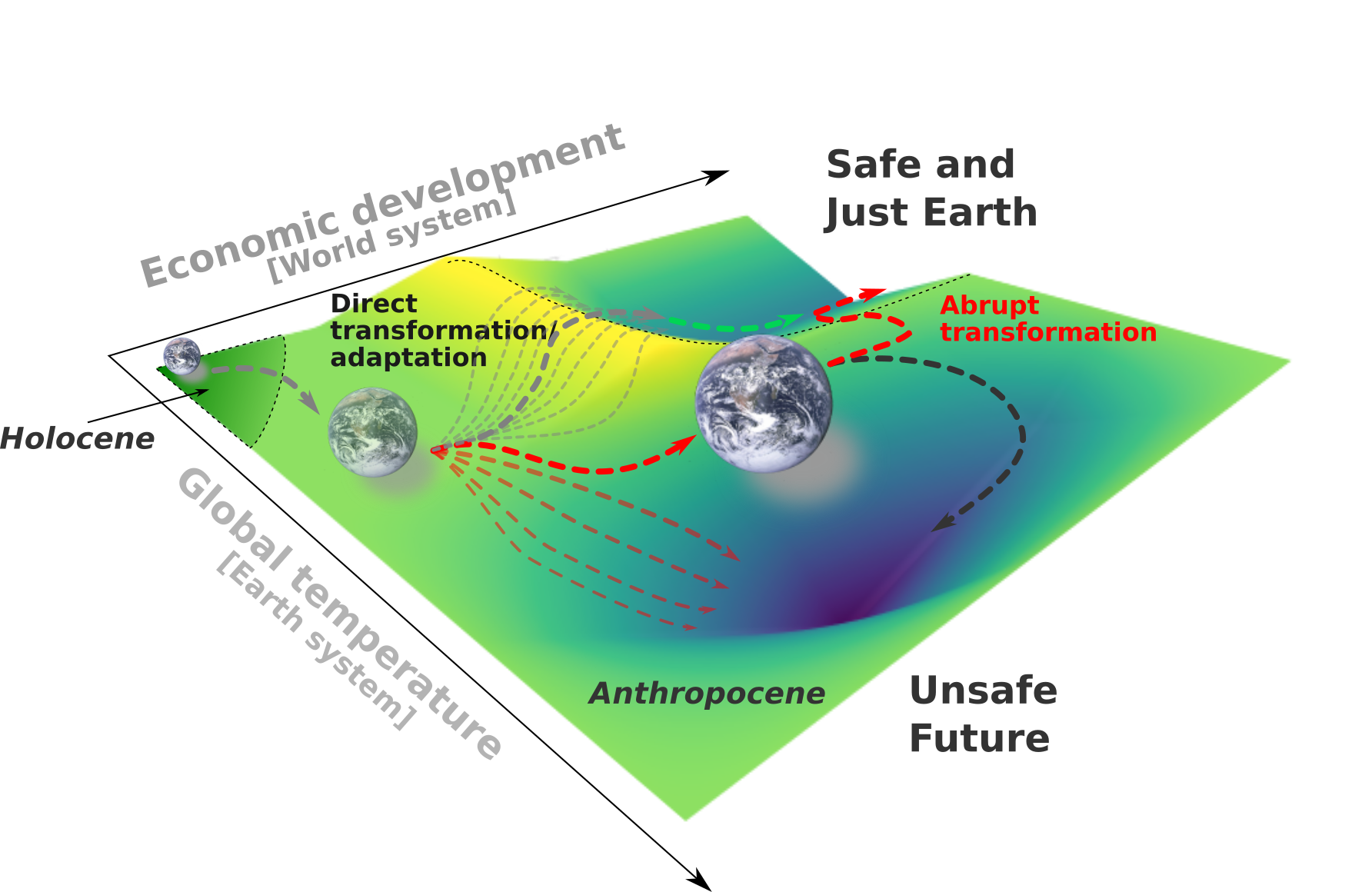}}
    \put(0,19){\includegraphics[width=0.45\textwidth]{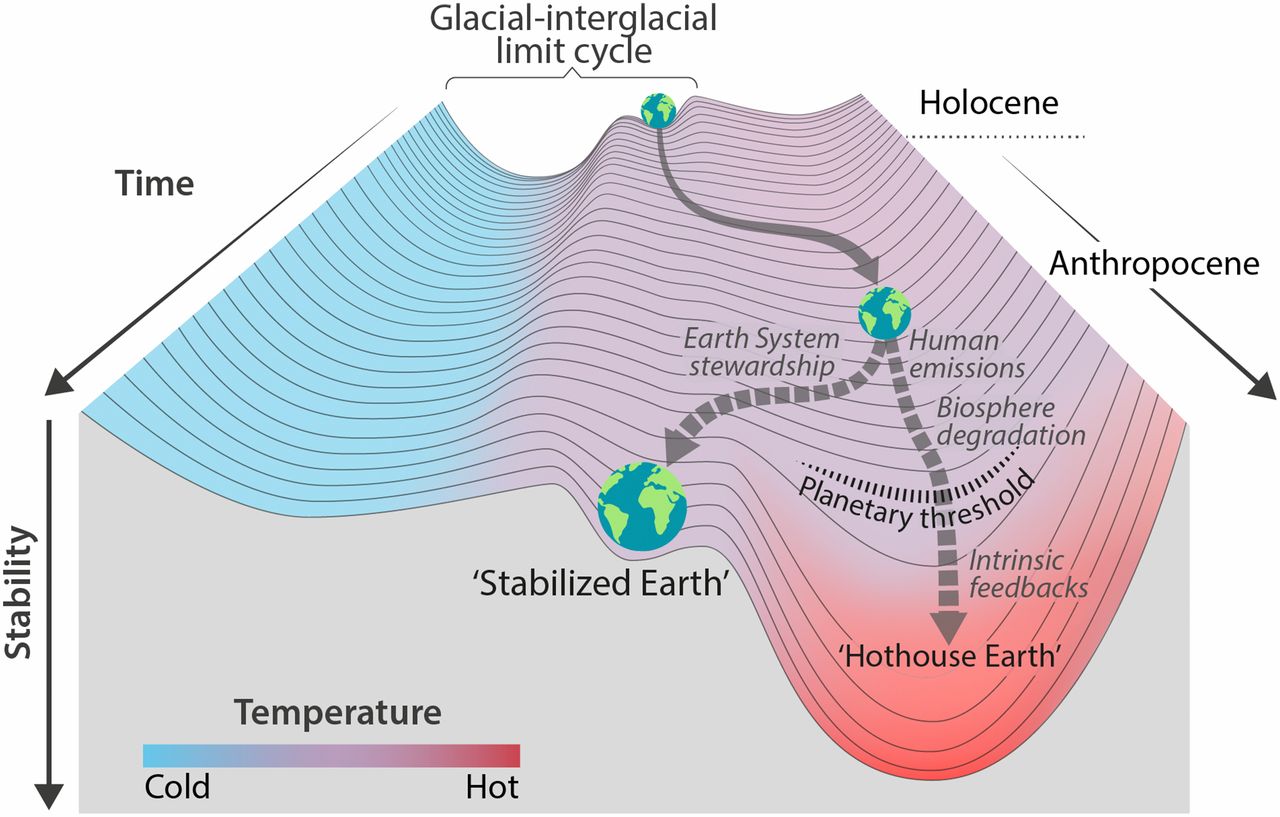}}
    \put(5,30){\parbox[t]{2in}{\footnotesize \sffamily World system \newline unlikely}}
    \put(16,19){\parbox[t]{2in}{\footnotesize \sffamily World system impossible}}
    \put(11.4,23){\parbox[t]{2in}{\footnotesize \sffamily World system \newline possible}}
    \put(0,37){\large \sffamily A}
    \put(0,16){\large \sffamily B}
   \end{picture} 
   
  \caption{\label{fig:WE_basins} A: Three basins of attraction following focused on biophysical aspects of the earth system over time. The possibility of sustaining a world system in each of these basins is indicated. B: Basins of attraction for the 'World-Earth System' where social/economic and biophyscial features interact. Earth enters the holocene near the zero economic development axis and has progressed to the center of the image now.  Humanity now faces critical decisions about navigating to the left (and how abruptly) or to the right.}

\end{figure}
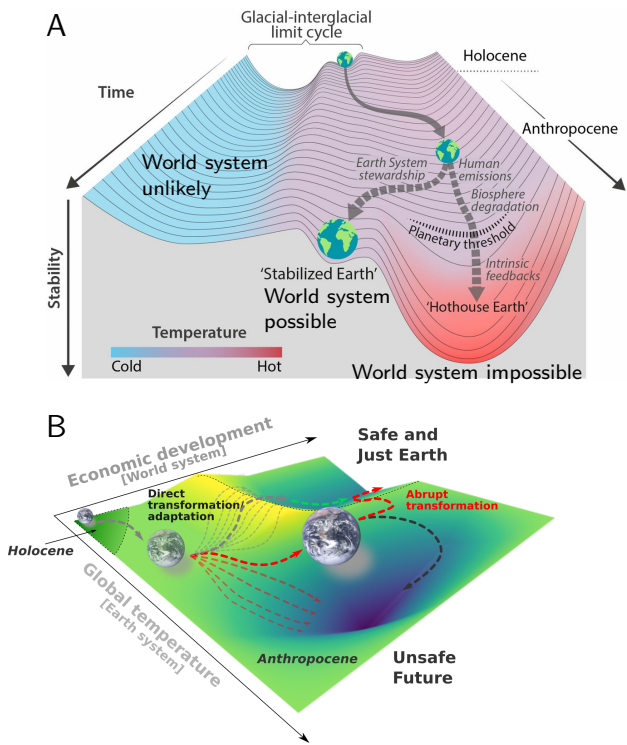

In the introduction, we identified three basins: interglacial, stabilized, and Hothouse Earth as characterized by Steffen et al. \citeyear{steffen2018trajectories}.  We have added the likelihood of sustaining a WS within each of these ES basins (Figure \ref{fig:WE_basins}A). Although a WS may be possible in the glacial-interglacial limit cycle, long periods of ice cover make this unlikely. Likewise, the environmental conditions in Hothouse Earth may be so severe as to make a WS impossible. 

Can nation states create a WS that is resilient to ES  and WS dynamcis? This is a challenging question.  We have a more modest goal:  in this paper we address the question of under what circumstances the WES can transition from one in which isolated regions support low-complexity societies to one in which interconnected regions support high-complexity societies. That is, what are the characteristics of the viable transition pathways between these two World--Earth states and how sensitive are these pathways to perturbations. In this sense we are taking a `pathway diversity' view of resilience \cite{lade2020resilience}, i.e.\ what is the resilience of the transformation pathway to exogenous shocks. An intuitive definition of WER at a particular state at a point in time is then the ratio of the number of paths that can reach an SJOS from that state to all paths than are accessible from that state. We will make this definition more precise in Section~\ref{sec:analysis}.

\section{The Model}

\begin{figure}[!ht]
  \newcommand{\xdim}{32}
  \newcommand{\ydim}{20} 
  \setlength{\unitlength}{0.1in}
  \centering
  \begin{picture}(\xdim,\ydim)
    \put(0,-1){\includegraphics[width=0.49\textwidth]{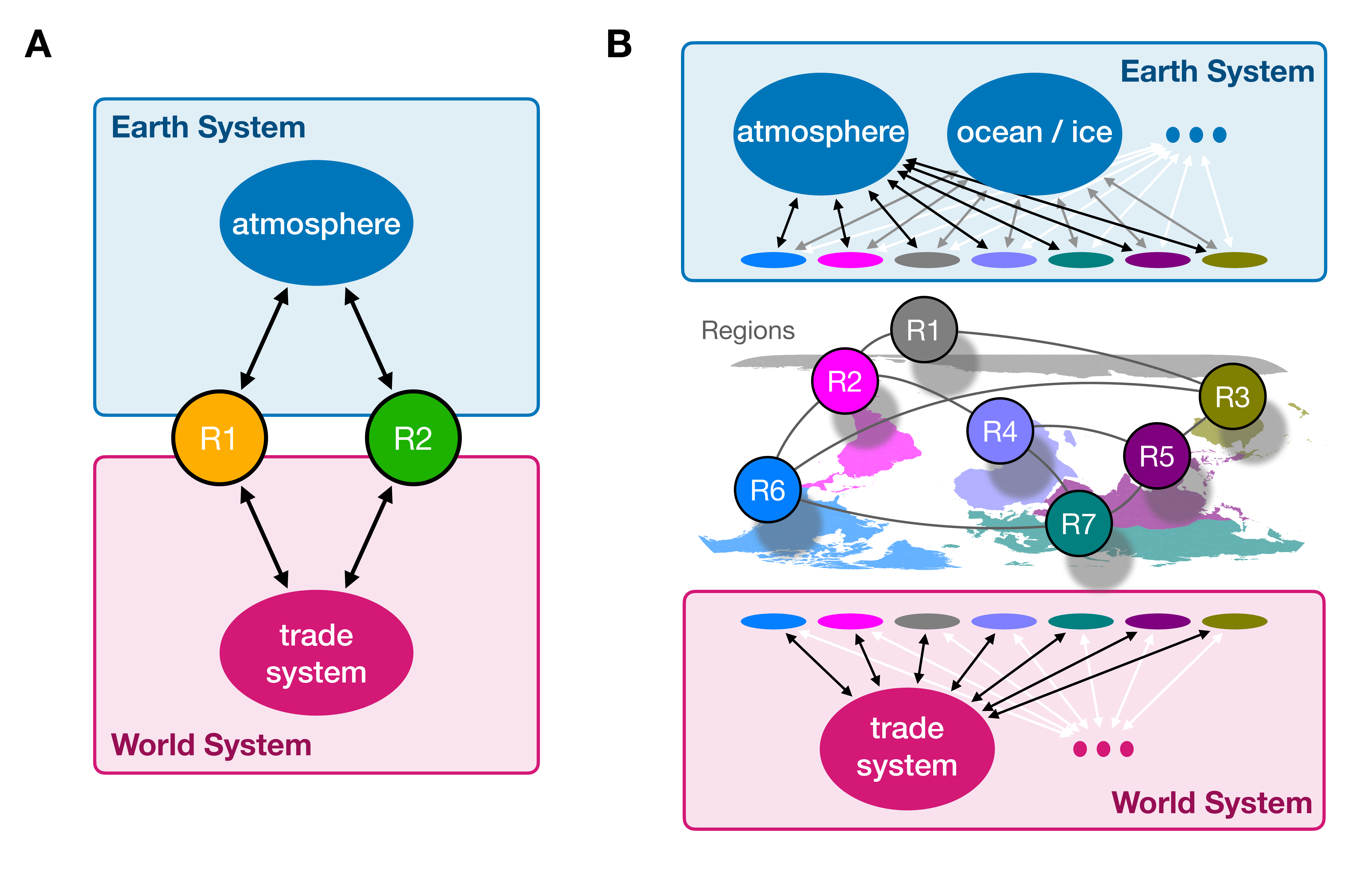}}
   \end{picture} 
   
\caption{\label{fig:WE_system} Schematics of World-Earth Systems....(A): 2-region system actually modelled. (B) spatial conceptualization of a 6 region (arbitrary, but maps roughly on to our system) including network representation of 6-region W-E system. }

\end{figure}

The notion of WES which we adopt as the fundamental unit of analysis is captured in Figure~\ref{fig:WE_system}.  Panel A shows the fundamental building block of any World--ES. On the bottom is the WS which connects, at a minimum, two polities supported by their respective natural capital endowments (R1, R2) through some sort of material or information exchange such as the example shown here of trade.  The regions (and the polities they support) are connected to the ES through material and energy exchanges through a shared entity of some sort (i.e. a common-pool resource) such as the example shown here of the atmosphere. WES are generated by networks of such building blocks as shown in Panel B where there are multiple possible connecting nodes in each, e.g. economic, human migration, cultural, and knowledge exchanges in the WS, and oceans, biodiversity, animal migration in the ES.  It is the endogenous dynamics generated by such networks that are the subject of this resilience analysis. We use this conceptualization to  extract the critical features of {\em a} WES of which {\em our} WES is a particular case. For clarity, we analyze a 2-region WES.  The essential feature of any world system are asymmetries in natural infrastructures, initial conditions and path dependencies (idiosyncratic shocks). These asymmetries generate differences in population dynamics (birth, death, and migration) and investment patterns across regions.  A 2-region model captures this essential feature of a WE system in the coarsest possible way and is thus where we start our analysis.

The model combines basic theory and empirical patterns from economic growth, population biology, and Earth-systems science.   The model tracks five state variables, two in each of regions 1 and 2: the human population size (billions of individuals), $P_1$ and $P_2$, the level of development of the built environment (capital-hours/year), $K_1$ and $K_2$, and one variable for the global system, the global externality (e.g. atmospheric carbon stocks) $G$. This leads to the dynamical system \\[-3mm]
\begin{eqnarray}
dP_j/dt &= & r_jP_j(1-P_j/\kappa_j) \\
dK_j/dt &=& I_j - \delta_j K_j - \sigma_j S_j(G) K_j\\
dG/dt  &= &e_1 Y_{i1} + e_2 Y_{i2} - uG + \theta (G-G_0)
\end{eqnarray}
for $j \in {1,2}$. The parameters $r_j$ and $\kappa_j$  represent the intrinsic growth rate carrying capacity of the human populations, $\delta_j$ the depreciation (entropic decay) of built infrastructure, and $\sigma_j$ the impact of shocks, $S_j(G)$ related to the global externality $G$ (e.g.\ storm surges, floods, etc. related to climate change).  The parameters $e_1$, $e_2$ and $u$ represent the carbon intensity of industrial production in regions 1 and 2 and the intrinsic assimilation capacity for G (e.g. carbon sequestration), respectively.  Finally, the function $\theta(G-G_0)$ represents a climate tipping point (threshold) about which we will say more later. 

The economic base of each region (GNP) consists of a `backstop' technology $Y_{bj}$ (e.g.\ simple agrarian economy) and  an `industrial' technology $Y_{ij}$.  Gross national income (GNI), $Y_{i}$, is then GNP plus the net of cross-region trade flows $Tr_{ij}$.  As industrial technology becomes more productive than the backstop technology, society transitions from the latter to the former thus capturing the `development' processes. By definition we have 
\begin{eqnarray}
Y_{1} =& Y_{b1} +  Y_{i1} - Tr_{12} + Tr_{21} \label{eq:gdp1}\\
Y_{2} =& Y_{b2} +  Y_{i2} + Tr_{12} - Tr_{21} \label{eq:gdp2}
\end{eqnarray}
where industrial production is modelled with standard Cobb--Douglas technology, i.e. $Y_{ij} = A_j K_j^{\alpha_j} L_j^{\beta_j}$ with $L_j$ and $A_j$ the labor supply (person-hours/year) and total factor productivity in region $j$,  and  with constant returns to scale ($\alpha_j + \beta_j = 1$).  Investment in each region is given by
\begin{equation}
    I_j = s_j \theta(Y_j - P_j C_m)(Y_j - P_j C_m)
\end{equation}
where $\theta(x)$ is a threshold function that is zero for  $x<0$ and rapidly increases to 1 for positive values of $x$.  This captures the idea that agents only invest after a minimum per-capita consumption level, $C_m$ is met and invest a proportion of this `disposable' income. Investment should be interpreted broadly to include hard (roads, canals, buildings) and soft (legal systems, information storage and transfer systems) infrastructures.

\begin{table*}[!tb]
  \footnotesize
  \begin{tabular}{cp{2in}cp{2.5in}} \hline
    Symbol
    &
      Defintion/units
    &
      Default value
    &
      Source
    \\ \hline

    $r_1,\,r_2$
    &
      Intrinsic growth rates (1/time)
    &
      0.038, 0.042
    &
      Fit to historical and projected UN data \newline \cite{UNpop2015}
    \\
    
    $\kappa_1,\,\kappa_2$
    &
      carrying capacity ($10^9$ persons)
    &
      1.5, 9.7
    &
      Fit to historical and projected UN data \newline \cite{UNpop2015}
    \\

    $\delta_{i}$
    &
      Entropic decay rates (1/time)
    &
      0.05
    &
      Standard value used in practice
    \\

    $\sigma_{i}$
    &
      Climate impact factor
    &
      0.03
    &
      Varied in the analysis
    \\

    $e_i$
    &
      Emmission intensity (hecto \newline ppm/$10^{12}$ \$)
    &
      0.0004
    &
      \cite{carbonIntensity,carbonIntensity2}
    \\

    $u$
    &
      Emmission uptake rate  (1/time)
    &
      0.0025
    &
     Based on historical estimated uptake of 280 pg over 200 years \cite{gruber2019oceanic}.
    \\

    $A_{1},\,A_2$
    &
      Total factor productivities
    &
      2.7, 1.7
    &
      Historical match of GNP \cite{worldGDP}.
    \\

      $\beta_1,\,\beta_2$
    &
      Output elasticity/factor share  of labor
    &
      0.5, 0.5
    &
      Varied in analysis. Historical range 1947-2016: 0.66-0.56. \cite{giandrea2017estimating}
      \\

  $\alpha_1,\,\alpha_2$
    &
      Output elasticity/factor share  of capital
    &
      0.5, 0.5
    &
      Varied in analysis. By assumption, $\alpha_i + \beta_i=1$ (constant returns to scale).
      \\
    
    $s_1,\,s_2$
    &
      National gross savings rates
    &
      0.25, 0.21
    &
      Varied in analysis. Rough historical average of around 0.25.  \cite{worldsavings}
      \\

  \end{tabular}
  \label{tab:parameters}
  \caption{Parameter definitions, units, and default values.  See text for definitions of other parameters that are varied extensively in the analysis.}
\end{table*}

\section{Analysis: Transition Pathways to SJOS}
\label{sec:analysis}

\begin{figure*}[!t]
  \centering
  \setlength{\unitlength}{0.1in} 
    \newcommand{\xdim}{65}  
    \newcommand{\ydim}{35} 
    \begin{picture}(\xdim,\ydim) 
      \put(0,0){\includegraphics[width=0.98\textwidth]{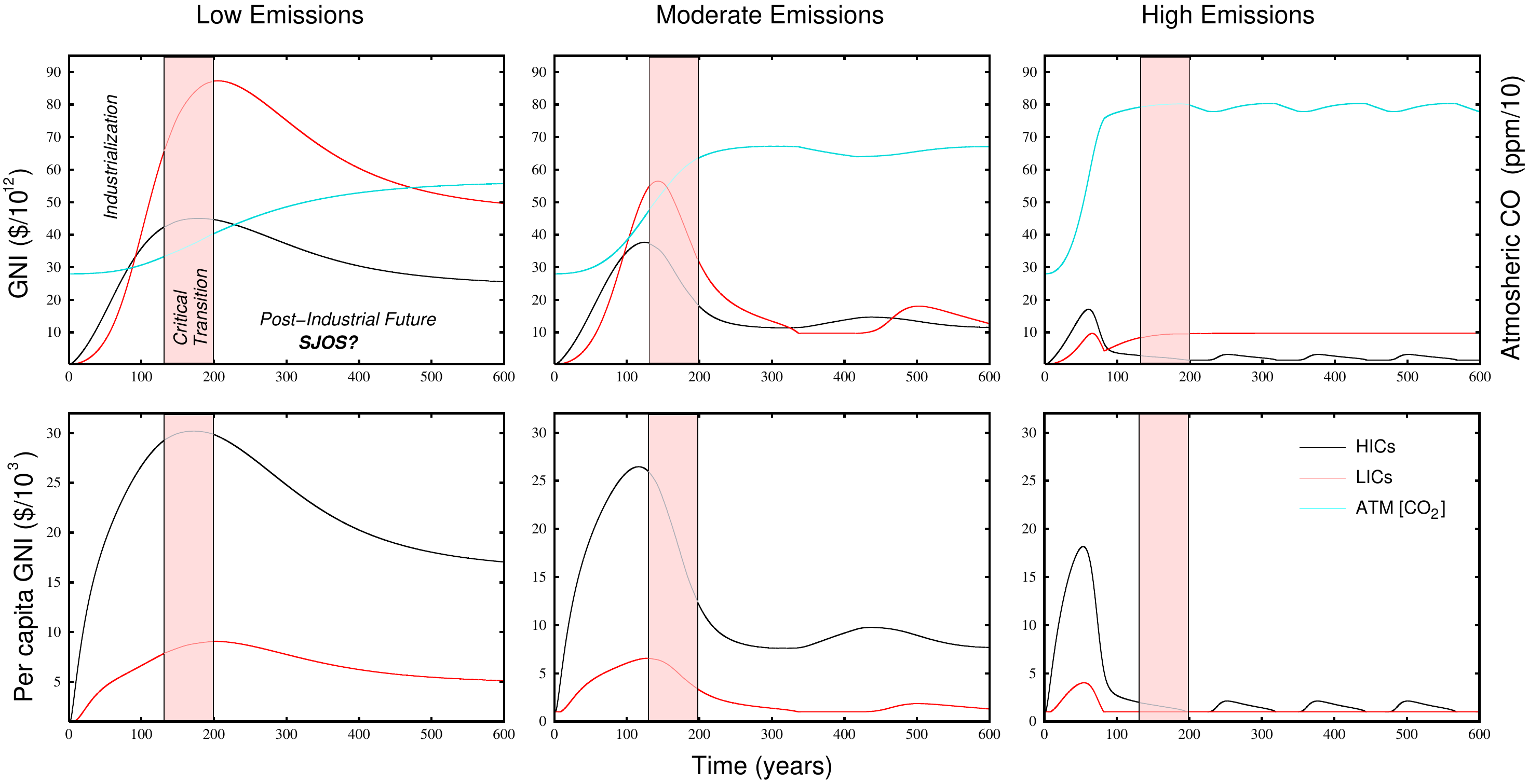}}
      \put(65.5,25.7){\sffamily \scriptsize \rotatebox{90}{2}}
    \end{picture} 
\caption{\label{fig:genBehave} General model behavior for low, moderate (roughly matching historical levels), and high emissions. These are trajectories for a 'friendly' earth systems that responds smoothly to increasing atmospheric carbon.  The black and red curves represent the "higher" and "lower" income countries. The colored strip represents a veil of uncertainty that separates the last 120 years of industrialization (high fossil fuel use and high pollution) from a post-industrial future through which societies must navigate.}
\end{figure*}

Our WER analysis focuses on how `tolerant' are the pathways available to us in the next critical transition phase (e.g.\ the remainder of this century) to our ignorance of the internal dynamics of the WE system.  If we hope to reach a SJOS, we must chart a robust path from our present unsafe, unjust, operating space (UUOS). Figure \ref{fig:genBehave} provides the baseline for our analysis.  These scenarios represent a `well-behaved' ES with no critical tipping elements.  The six panels show representative trajectories for two regions.  The two regions --- ``high-income countries'' (HICs) and ``low-income countries'' (LICs) --- differ only in their savings rates and total factor productivities. This difference may be due to different cultural contexts (less propensity to save) and organizational capacities (variation in contract creation and enforcement across regions) and roughly matches experiences in the twentieth century. The colored strip represents a critical transition period through which we must find resilient pathways to reach a SJOS in the post-industrial future.  Using the Spaceship Earth metaphor, we must gain enough knowledge of the functioning of the ship's life support systems and transform the social system on board to enable the life-support system to be maintained indefinitely in the future, and whatever solution we come up with, cannot be too fragile.  

The three scenarios are identical except in their per-unit-of-GDP carbon emissions.   The scenarios begin with the situation in 1900 with the take-off of fossil-fuel-based industrialization and atmospheric CO$_2$ concentration (ACC) of around 280\,ppm.  The critical transition phase corresponds to the calendar year of 2020 and lasts to the end of this century. In the low emissions case, ACC reaches around 320\,ppm in 2020. There is a soft landing in the post industrial future but as the bottom panel shows, development remains significantly unequal.  It is important to note that this unequal development enables the HICs to reach such a high per-capita standard of living (bottom panel, column 1) before it begins to decline.  

The moderate emissions case roughly maps onto Earth's historical trajectory.  In 2020, ACC is around 420, total GDP of HICs is around 35 trillion and LICs is around 55 trillion for a total of around 90 trillion current USD \cite{carbonIntensity}. In this case, ACC tops out at just under 700 at century's end.  This concentration induces significant costs and by the end of the century, world output drops to around 50 trillion and, in the long run declines to around 30 trillion with HICs and LICs contributing about equally.  This masks the per-capita story with the HIC's able to maintaining industrial economic structures with 10 times per-capita GNI of LICs. LICS revert back to baseline economic structures (more agrarian-based with an per-capita income of around 1000 USD).  

\begin{figure}[!t]
  \centering
  \includegraphics[width=0.48\textwidth]{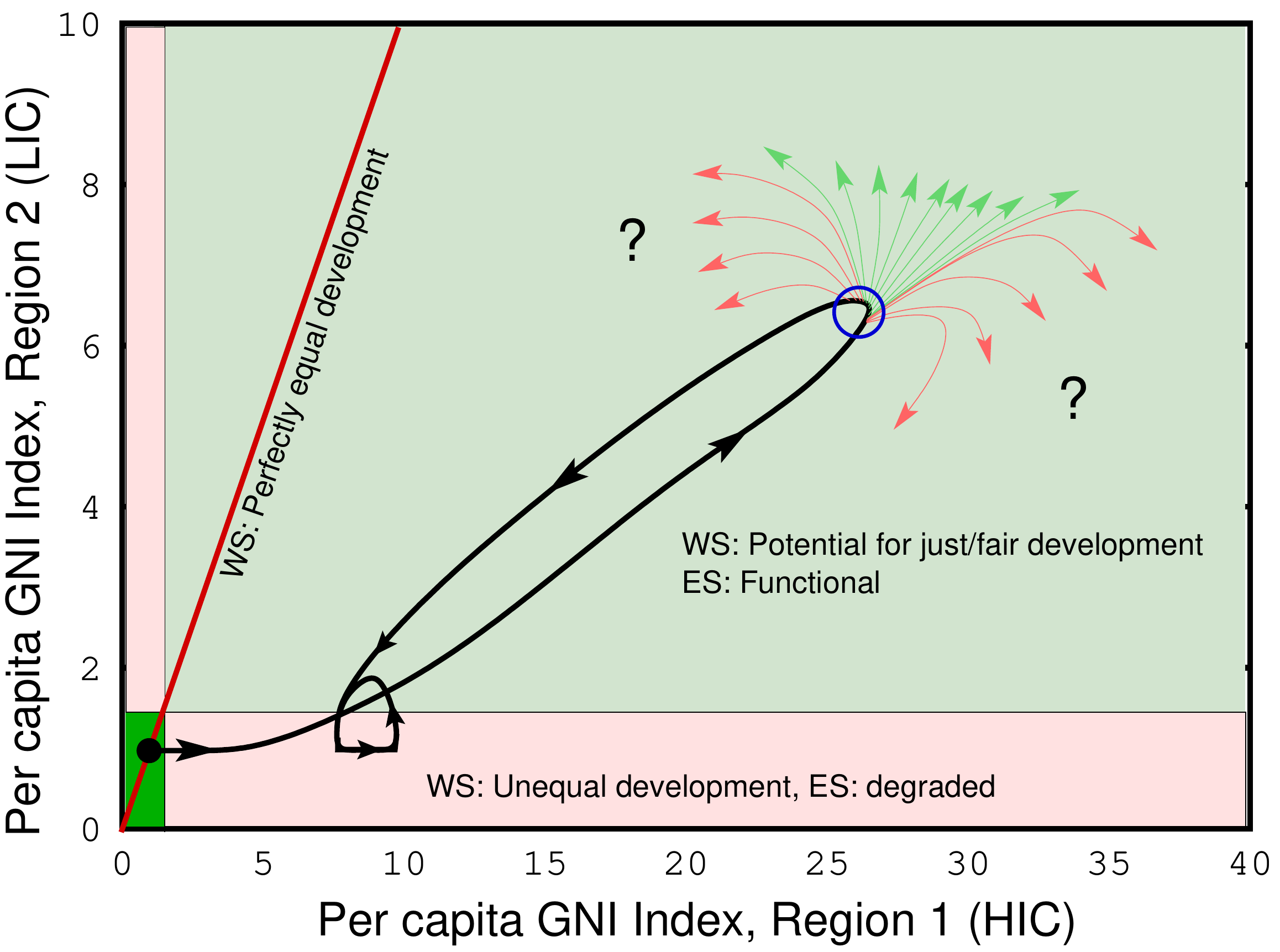}
\caption{\label{fig:gniPhasePlane} Phase plane digram for per capita GNI in higher and lower income counties. The phase plane represents in rigorous mathematical terms the methaphorical landscape depicted in Figure~\ref{fig:WE_basins}.}

\end{figure}

Figure \ref{fig:gniPhasePlane} shows the moderate emissions case in per-capita GNI phase space.  The heavy black curve is the business as usual scenario and illustrates three possible attractors. The red, 45 degree line indicates equal per capita GNI across regions. The dark green region in the lower left is the pre-industrial, baseline technology region where the impact of humans on the ES is minimal (most of the Holocene).  The pink regions indicate a degraded ES wherein emissions associated with industrial production are very high (top center panel, Figure \ref{fig:genBehave}).  The limit cycle on the lower left results from the ES pulling carbon out of the atmosphere which reduces economic impacts just enough for a period of mild economic growth which then increases atmospheric carbon again. The blue circle at the maximum development point represents a window of opportunity for decarbonization and/or more balanced growth across regions. The trajectory ensemble (or bundle) illustrates path-based resilience. Green paths enable continued development. Salmon paths lead to attractors with unequal development and/or a degraded ES. These paths may be the result of increasingly unequal development that delays or prevents action on decarbonization policies due to inter-regional disagreements (right) or insufficient decarbonization policy in spite of efforts to promote more equitable development across regions (left).  WER can then be computed as the `volume' of the state space containing the green trajectories (SOS) relative to that containing the salmon trajectories and the probability of being knocked out the SOS.  

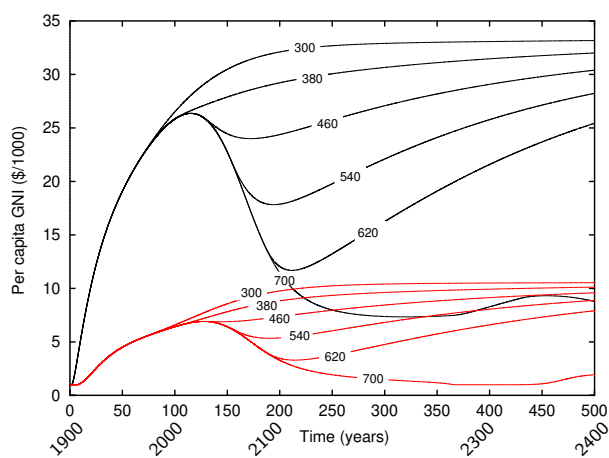
\begin{figure}[!ht]
\centering
\setlength{\unitlength}{0.1in}
\begin{picture}(33,24)
\put(0,0){\includegraphics[width=0.47\textwidth]{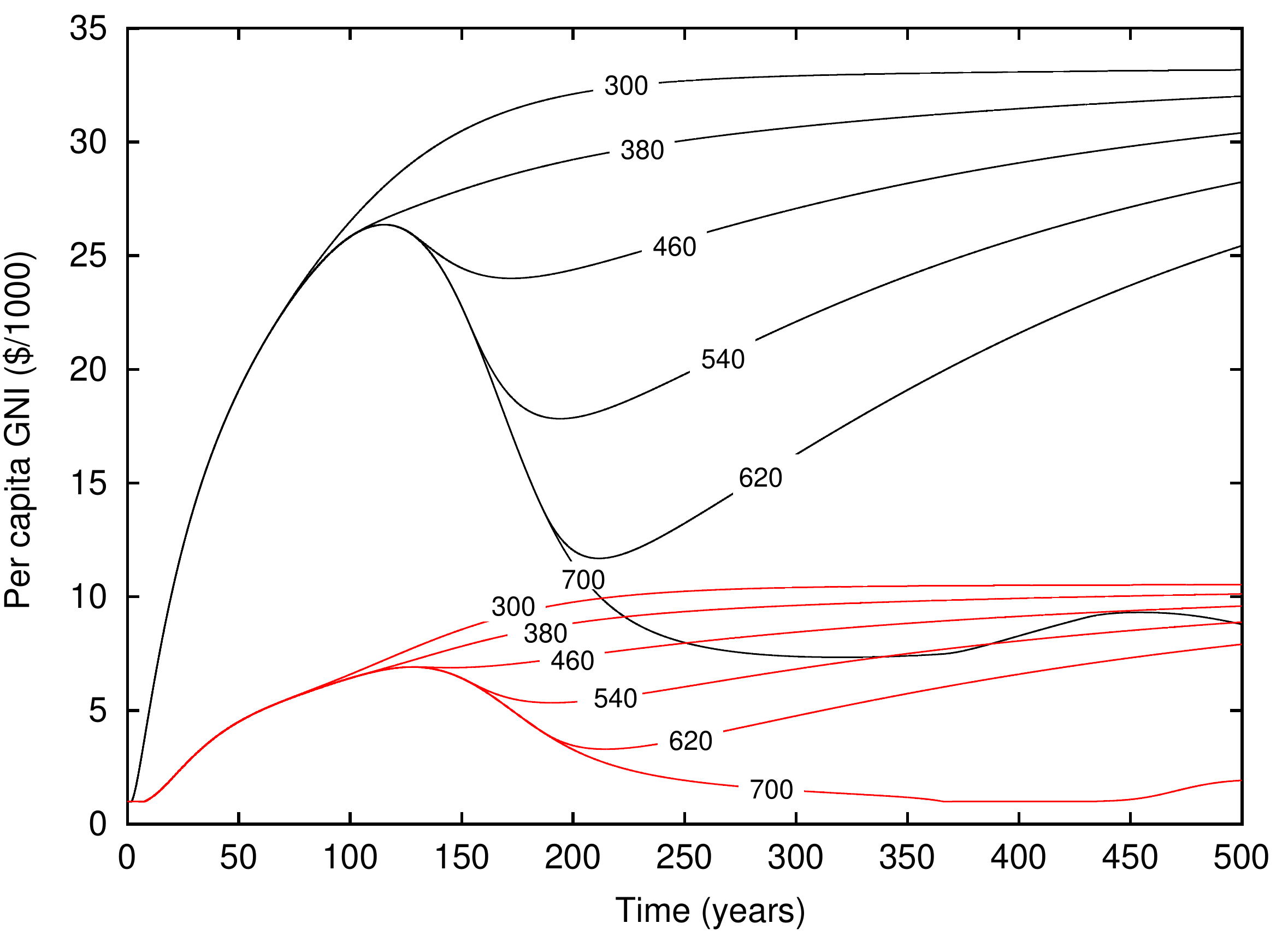}}
\put(2,-0.5){\rotatebox{45}{\sffamily \scriptsize 1900}}
\put(7.2,-0.5){\rotatebox{45}{\sffamily \scriptsize 2000}}
\put(12.4,-0.5){\rotatebox{45}{\sffamily \scriptsize 2100}}
\put(23.5,-0.5){\rotatebox{45}{\sffamily \scriptsize 2300}}
\put(29.5,-0.5){\rotatebox{45}{\sffamily \scriptsize 2400}}
\end{picture}
\caption{\label{fig:decarb_nice} Decarbonization scenarios when there are no  tipping points in the earth system. The black and red curves refer to HIC and LIC respectively.  The numbers refer to the atmospheric carbon concentration at which the decarbonization project that reduces carbon concentration by 10\% annually begins. Plausible calendar dates for this carbon release process for our WE system are shown at angles on the x-axis.}
\end{figure}

Figure \ref{fig:decarb_nice} shows time trajectories of per capita GNI with various decarbonization programs initiated at different ACC levels ranging from 300-700\,ppm and proceeding at 10\% per year reduction (roughly 90\% decarbonization in 20 years and complete decarbonization over 40-50 years). In this case, the ES is `friendly', i.e., there are no thresholds. The main message is that unequal development between regions enables HICs to achieve much higher levels of development than would have been possible had LICs used up more of the (open access) waste assimilation capacity of the environment and, as a result, failure to decarbonize will be more painful for HICs. Thus, there is a strong incentive for HICs to lead on decarbonization, act early, and convince LICs to join them. 

\begin{figure*}[!ht]
  \centering
    \includegraphics[width=0.48\textwidth]{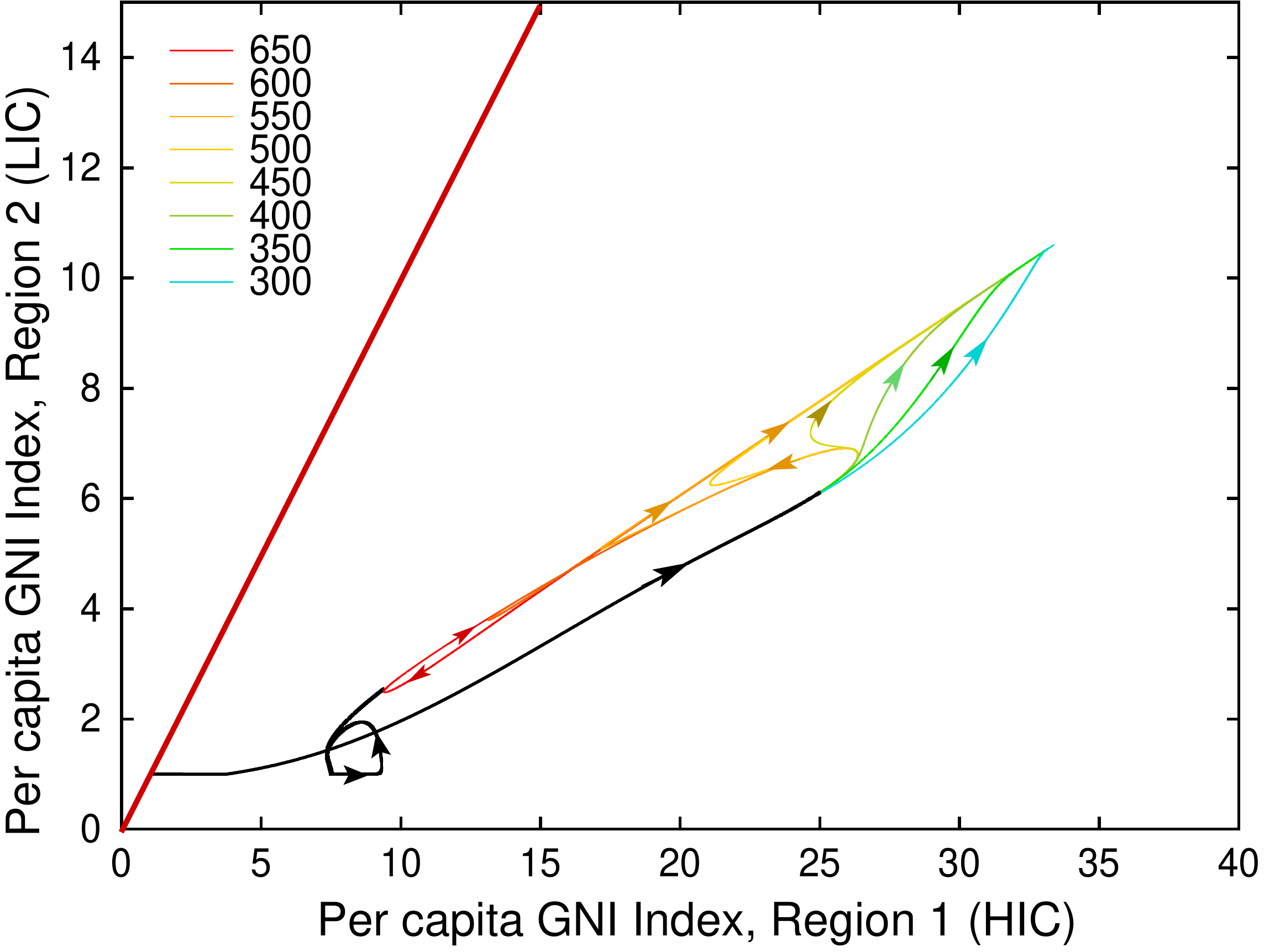}
    \includegraphics[width=0.45\textwidth]{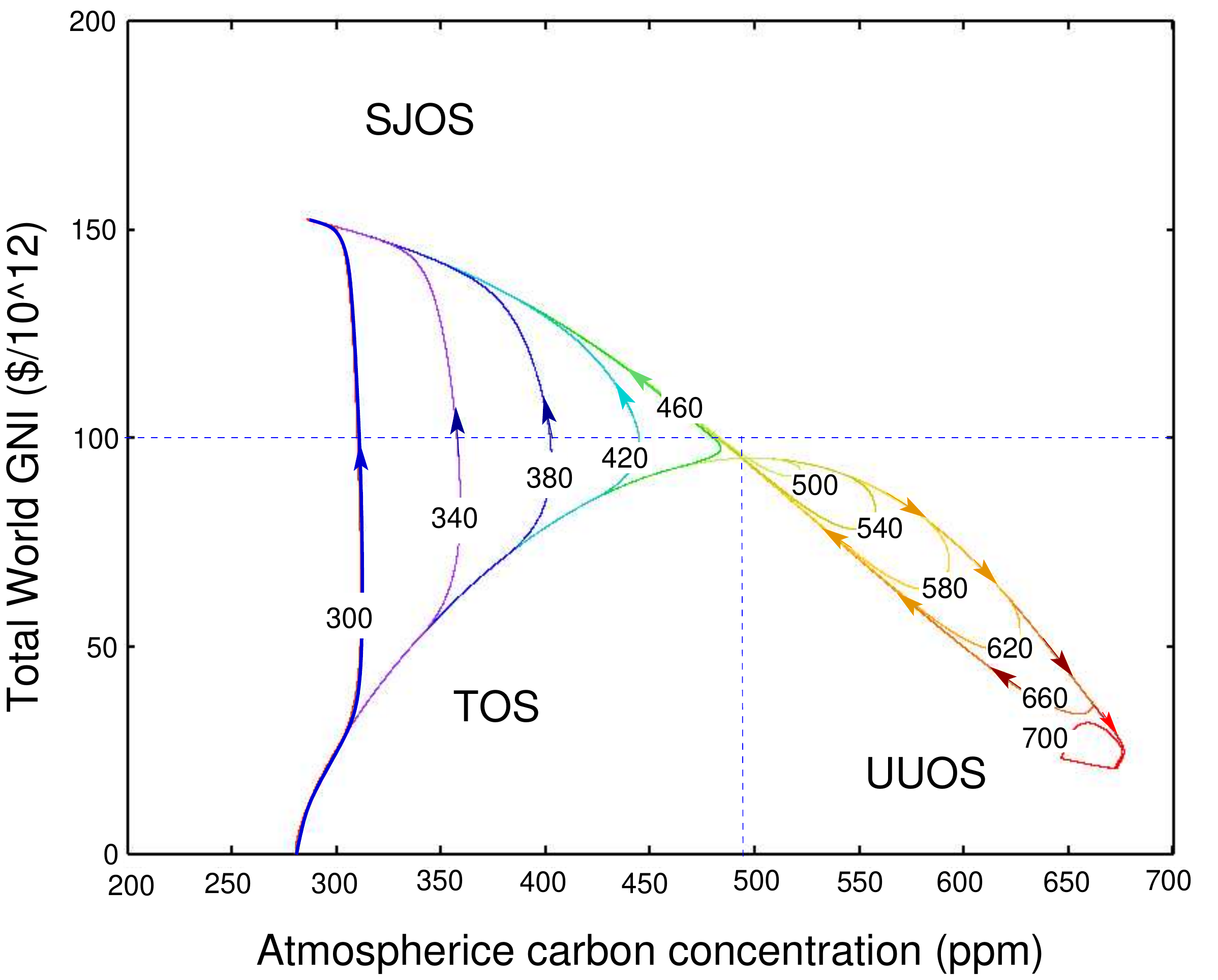}
\caption{\label{fig:decarb_nice_pp} Phase plane representation of  scenarios in Figure \ref{fig:decarb_nice}, i.e. per capita GNI for region 2 versus region 1 in trillions of dollars.  On the right is the total GNI versus atmospheric carbon. Note, the color coding doesn't match exactly. TOS= transitional operating space, SJOS = safe and just operating space, UUOS = unsafe, unjust operating space. See text for further discussion.}
\end{figure*}

Figure \ref{fig:decarb_nice_pp} shows the same information from Figure \ref{fig:decarb_nice} in the phase plane.  Note that with no climate tipping elements, the system can always recover from the business-as-usual trajectory (black) - it is simply a matter of how much economic disruption the system experiences.  There are two important takeaways from this figure: decarbonization alone reduces inequality (development paths tend closer to equal development path) and the dividing line between paths that involve some level of economic disruption (turn right and down before turning up and left) is somewhere between 400 and 450\,ppm for the historically calibrated parameter set.

In the right panel is the mathematically rigorous phase space representation of the metaphor in Figure  \ref{fig:WE_basins}B.  Notice that in the `nice' world, the system can enter and remain in the UUOS for some time until decarbonization can pull it out.  In this case, again referring to Figure \ref{fig:WE_basins}, the depth of the unsafe basin is determined completely by social factors.  The WE system can always be pulled out of this basin by social and economic processes if WS resilience is high due to its capacity to adapt to decreasing levels of wellbeing. In this case, navigating the landscape between the SJOS and UUOS is not necessarily perilous, i.e. pathway resilience is high because the ES resilience is high.  This analysis also illustrates why the ball-and-cup visualization breaks down in higher dimensional models--we need to think in terms of bundles of paths.

\begin{figure*}[!ht]
\centering  
\includegraphics[width=0.4\textwidth]{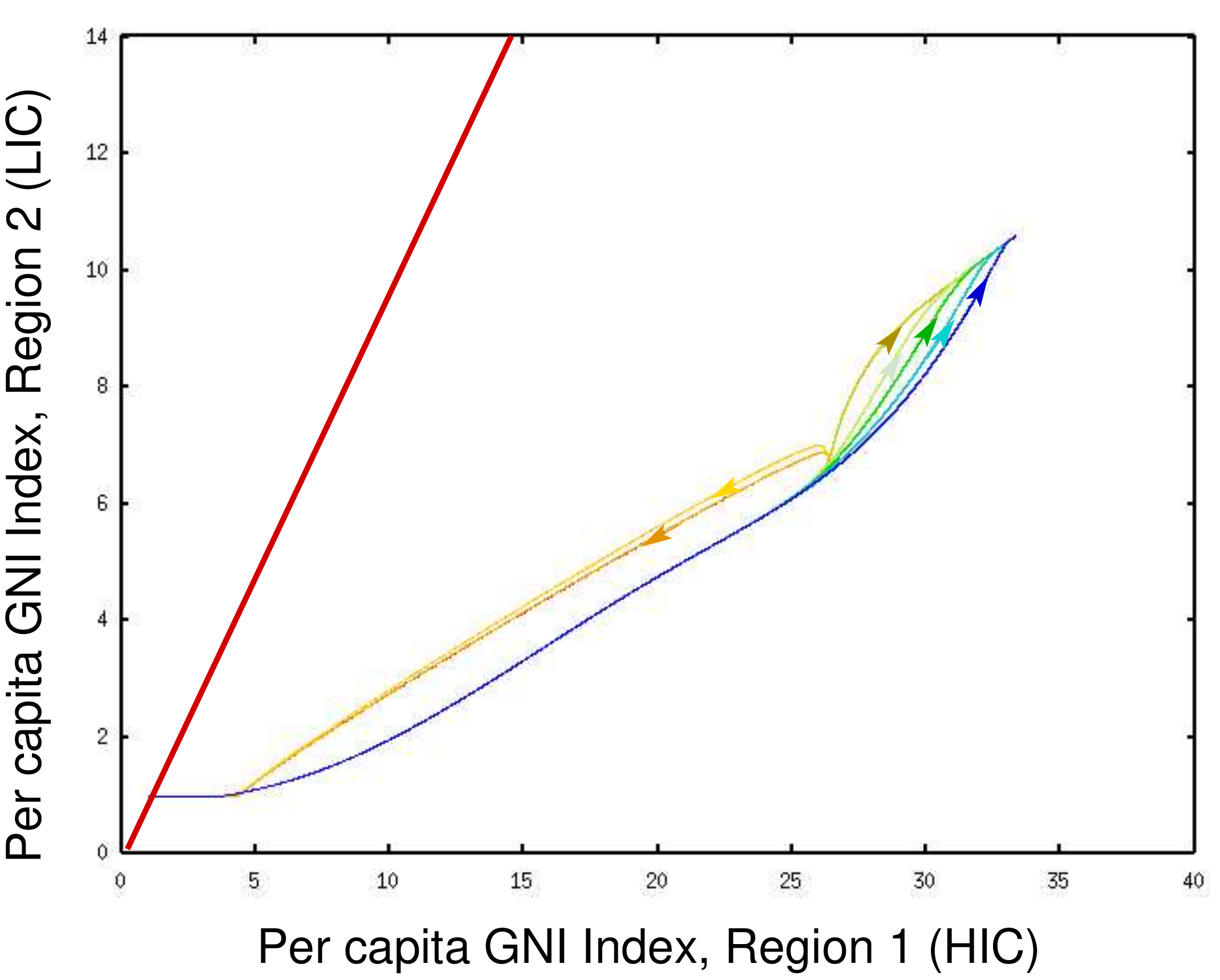}
\includegraphics[width=0.45\textwidth]{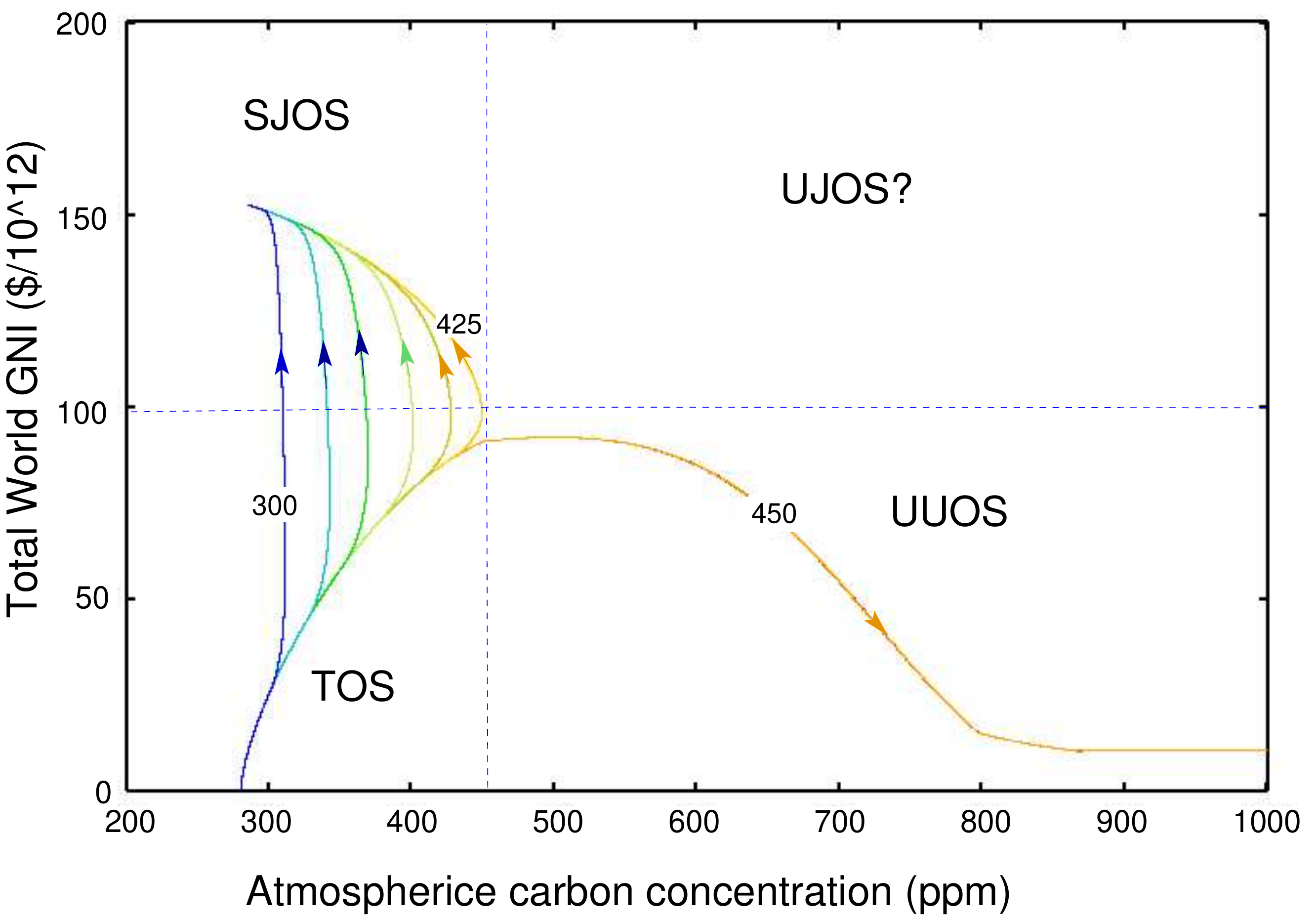}
\caption{\label{fig:decarb_not_nice_pp} Phase plane representation of decarbonization scenarios with a climate threshold at 450 ppm. Per capita GNI is trillions of dollars. As the colors range from blue to yellow orange, the decarbonization threshold ranges from 300 ppm to 450 ppm, respectively.  The critical threshold value occurs at approximately 425 ppm. Below 425 ppm, trajectories go to the SJOS. Otherwise they go to the UUOS.}
\end{figure*}

Figure \ref{fig:decarb_not_nice_pp} is the analogue of Figure \ref{fig:decarb_nice_pp} with a climate threshold at 450\,ppm.  Now there is a stark division between trajectories trapped in the UUOS and those that can reach the SJOS.  Those that reach the SJOS initiate decarbonization just under approximately 425 ppm. Policy action must apply the breaks 25 ppm before the threshold to compensate for system inertia.  The `ridge' has collapsed to a razor's edge.  In this case, the threshold on the global externality creates a basin in the UUOS.  It is important to note that thresholds in the social system could generate this situation as well.  In Figure \ref{fig:decarb_nice_pp}, the trajectories are distinguished only by {\em when} decarbonization occurs not {\em whether} it occurs. It is reasonable to assume that when economic growth and well-being start to decrease globally because of climate damages (e.g. around 490 ppm in Figure \ref{fig:decarb_nice_pp}), there may be a point beyond which the appetite for contributing to the public good of decarbonizing goes to zero. This wealth-dependency of contributions to public goods \cite{heap2016endowment} reduces the resilience of the world sytem to inequality and thus can generate the same type of lock-in effect as loss of resilience in the ES.

As with any even relatively low-dimensional model, there is a large number of parameter choices we could make and with them, a large number of scenarios. For example
\begin{itemize}
    \item Will geoengineering remove climate thresholds and thus transform Figure~\ref{fig:decarb_not_nice_pp} into Figure~\ref{fig:decarb_nice_pp}? In our model, this translates roughly into the next question - i.e. how fast can you draw down atmospheric carbon.
    
    \item Where is the climate threshold? For example, if the climate threshold is at 500 ppm, society must start decarbonizing at 10\% per year when the ACC is 476 ppm. This is slightly closer to the threshold than with the 450 ppm threshold (24 versus 25 ppm less than the threshold) likely due to the fact that at around 470 ppm, climate damages have begun to bite harder than at 420 ppm so economic inertia will be slightly less allowing society to hit the breaks a split second (1 ppm) later.  
    
    \item Will the regions have differential impacts due to G?  If for example, the impact of G on Region 2 is higher than Region 1, the bundle of the trajectories in Figures \ref{fig:decarb_nice_pp} and \ref{fig:decarb_not_nice_pp} would rotate clockwise.  In the case without a climate  threshold, the long run attractor shifts to the right and the limit cycle vanishes. In this extremely unjust outcome, Region 2 can never recover from climate damages even temporarily and Region 1 benefits in that its long run economic output is slightly higher.  Thus, Region 1 contributed to G more historically thus bearing more responsibility for creating the problem, and Region 2 bears the costs disproportionately. 

\end{itemize}
These issues do not change the basic qualitative dynamics summarized in our analysis - they may shift the trajectories in state space but won't change the underlying topology of the state space. This is why we have emphasized that we are not attempting to capture {\em the} WE system we live in but, rather {\em a} WE system that has the same fundamental features as {\em any} WE system including the one we live in.

What is clear, however, is that less developed countries' (LIC in our model) willingness to decarbonize hinges on inequality and the willingness of rich countries to provide aid.  At COP 26, India demanded that rich nations pay 1 trillion USD before it would make a climate pledge \cite{straittimes2021,bloomberg2021} and demanded that rich countries acknowledge their historic responsibility \cite{guardian2021}.  Four countries, Brazil, China, India, and South Africa joined forces to tie emission cuts to funding from wealthy countries \cite{riotimes2021}. Thus, the question of how 
economic inequality may affect decarbonization seems more immediate and important than long-debated biophysical details.

Trajectories in Figure \ref{fig:decarb_not_nice_pp} assume that both HICs and LICs decarbonize starting at the same time and at a rate of 10\% per year. The recent news mentioned above suggests this is unlikely.  To explore the implications of inequality, we suppose that $re_2 = re_1 (1-\lambda_e + \lambda_e \cdot pcGNI_2/pcGNI_1)$ where $re_i$ is the decarbonization rate of region $i$ and $\lambda_e \in [0,1]$ is the weight region 2 (LIC) puts on income inequality in choosing its decarbonization rate. If $\lambda_e = 0$, region 2 matches region 1 in their decarbonization rate (the scenario in Figure \ref{fig:decarb_not_nice_pp}).  If $\lambda_e = 1$, region 2 decarbonizes at a proportion of region 1's rate given by the ratio of its per capita GNI to that of region 1.  In this case, the threshold to act is at 390 ppm, not 425!  In an equal world, we may have had a change to act now.  If ineqaulity matters as news reports suggest, we have already missed our opportunity to act or, rather, we need to deal with inequality very quickly.  For example, to increase the threshold back to 425 by directly reducing inequality would have required that the HICs transferred 45\% of their GDP to LICs  ($Tr_{12} = 0.45 Y_1$ and $Tr_{21}=0$ in (\ref{eq:gdp1}) and (\ref{eq:gdp2})) from the start of the industrial revolution!  Of course, this is not remotely tenable or fair in that the HICs would have had to economically subsidize populations in LICs much larger than their own to a per-capita income level in the HICs.  Given that such enormous historical wealth transfers are untenable, the only hope is that the gesture of providing aid will reduce {\em perceived} unfairness and reduce $\lambda_e$.  That is, the model suggests that one of the most important things the HICs can do is to work with LICs to reduce 
$\lambda_e$.

\section{Discussion/Conclusions}

We have developed a framework within which to conceptualize WER.  We have emphasized that the basin of attraction notion is not particularly useful in analyzing WER because we are not in a basin we can stay in. We are on a trajectory to a new basin and we have to avoid falling into very undersirable basins.  Thus, we must think in terms of pathway resilience, i.e. the relative number of paths that allow us to move from the TOS to the SJOS. We then developed a mathematical model to formalize this conceptualization and demonstrated how in a resilient ES (Figure \ref{fig:decarb_nice_pp}) pathwise resilience depends soley on a resilient world system capable of acting to decarbonize.  In a less resilient ES, many paths between the TOS and SJOS vanish (Figure \ref{fig:decarb_not_nice_pp}).  This dramatic loss of potential pathways illustrates the critical importance and value of investing in ES resilience.  Our findings show that ES resilience is probably our only chance to reach the SJOS.

Next we illustrated the importance of WS dynamics by showing how the introduction of the notion of fairness coupled with regional inequality further restricted the number of viable paths from the TOS to the SJOS. In some broad sense, inequality reduces the resilience of the WS and, with it, the resilience of the WES. This model outcome is consistent with climate change negotiations historically.  

Specifically, prior to the UN Climate Change Conference in Paris, the focus was on `contraction and convergence'.  Based on historical responsibility, HICs recognized  that LICs should be allowed to decarbonize more slowly to compensate for economic hardship.  However, evidence of the rapid decline of ES resilience at the Paris meeting prompted the realization that `contraction and convergence' was not tenable. Rather HICs and LICs had to both decarbonize more quickly and at the same rate. The only way to facilitate this was through a  compensation scheme whereby HICs provided funds to LICs for decarbonizaton. The problem, as with all public goods, is under-provision: the Green Climate Fund is likely woefully inadequate for the task. The continued importance of compensation was evident at the Glascow Climate Meeting \cite{riotimes2021}.  Based on the model, if we include justice considerations, we are well past the point (390 ppm) of acting to avoid crossing a 450 ppm threshold.  Given that it is impossible to address historical inequalities now, we must rely on the goodwill of LICs to reduce $\lambda_e$.  HICs must set and meet reasonable contribution levels to the Green Climate Fund to earn this goodwill. 

The analysis here takes just a first small step in analyzing WER. First, we have neglected some key processes such as migration and technological change. Second, and much more importantly, we analyzed the system {\em as if} we had perfect information about the system.  If this were the case, resilience is a moot point.  We can simply calculate the location of boundaries and decide what to do. We need not invest in resilience per se. However, the reality is that we don't have perfect information about  the system.  To properly assess resilience will require a careful Monte Carlo analysis of our model yielding an assessment of the probability that we may successfully navigate Spaceship Earth from the TOS to the SJOS. This probability will provide a combined measure of the state of our spaceship's life support system (ES resilience) and the capacity of the crew and passengers to act decisively (world system resilience), i.e. of WER. Computing this measure is the focus of future work using this model.

\section*{Acknowledgements}

We are grateful for financial support by the European Research Council Advanced Grant project ERA (Earth Resilience in the Anthropocene, grant ERC-2016-ADG-743080) and the Leibniz Association (project DominoES). The authors thank R. Winkelmann and the participants of the LOOPS-4 workshop on Earth resilience in the Anthropocene held in Bad Belzig, Germany in March 2019 for insightful discussions.

\section*{References}

\bibliographystyle{jphysicsB}
\bibliography{bibliography}

\end{document}